\documentclass[pre,aps,twocolumn,superscriptaddress,notitlepage]{revtex4-1}
\usepackage[colorlinks,linkcolor=darkblue,citecolor=darkblue,urlcolor=darkblue]{hyperref}
\usepackage{amsmath}
\usepackage{amsfonts}
\usepackage{xcolor}
\definecolor{darkblue}{rgb}{0,0,0.6}
\usepackage{verbatim}
\usepackage{graphicx}
\usepackage{esint}
\usepackage{float}
\usepackage{bm}
\usepackage{bbm}
\usepackage{circledsteps}
\usepackage{tikz,standalone}
\usetikzlibrary{matrix,positioning}
\newcommand{\beq}{\begin{equation}}
\newcommand{\eeq}{\end{equation}}
\newcommand{\bfr}{\boldsymbol{r}}
\newcommand{\bfp}{\boldsymbol{p}}
\usepackage{soul}

\begin{document}

\title{Intermittent relaxation and avalanches in extremely persistent active matter}

\author{Yann-Edwin Keta}

\affiliation{Laboratoire Charles Coulomb (L2C), Universit\'e de Montpellier, CNRS, 34095 Montpellier, France}

\newcommand{\gaug}
{\affiliation{Institute for Theoretical Physics, Georg-August-Universit\"at G\"ottingen, 37077 G\"ottingen, Germany}}

\author{Rituparno Mandal}
\gaug

\author{Peter Sollich}
\gaug

\affiliation{Department of Mathematics, King's College London, London WC2R 2LS, UK}

\author{Robert L. Jack}

\affiliation{Yusuf Hamied Department of Chemistry, University of Cambridge, Lensfield Road, Cambridge CB2 1EW, United Kingdom}

\affiliation{Department of Applied Mathematics and Theoretical Physics, University of Cambridge, Wilberforce Road, Cambridge CB3 0WA, United Kingdom}

\author{Ludovic Berthier}

\affiliation{Laboratoire Charles Coulomb (L2C), Universit\'e de Montpellier, CNRS, 34095 Montpellier, France}

\affiliation{Yusuf Hamied Department of Chemistry, University of Cambridge, Lensfield Road, Cambridge CB2 1EW, United Kingdom}

\date{\today}

\begin{abstract}
  We use numerical simulations to study the dynamics of dense assemblies of self-propelled particles in the limit of extremely large, but finite, persistence times. In this limit, the system evolves intermittently between mechanical equilibria where active forces balance interparticle interactions. We develop an efficient numerical strategy allowing us to resolve the statistical properties of elastic and plastic relaxation events caused by activity-driven fluctuations. The system relaxes via a succession of scale-free elastic events and broadly distributed plastic events that both depend on the system size. Correlations between plastic events lead to emergent dynamic facilitation and heterogeneous relaxation dynamics. Our results show that dynamical behaviour in extremely persistent active systems is qualitatively similar to that of sheared amorphous solids, yet with some important differences. 
\end{abstract}

\maketitle

\section{Introduction}

Active matter systems continue to surprise and challenge us with their range of diverse behaviour, generating new insights into systems that are relevant for both biology and fundamental statistical mechanics~\cite{vicsek95,marchetti13,volpe16,solon15}. In dense disordered systems with slow dynamics, systems of active particles exhibit glassy behaviour reminiscent of passive equilibrium systems, as well as interesting new phenomena~\cite{berthierkurchan13,berthier14,ni13,szamel2015glassy,mandal16,leomach19,berthier2019glassy,janssen2019active,mandal2020extreme,ManSol21,keta2022disordered}.

A fascinating idea is that collective motion observed on large length scales may be triggered by driving forces acting at the particle scale.  This represents a form of dynamic self-organisation that emerges out of equilibrium from many-body effects. A wide variety of collective behaviour is observed in active systems, from bacterial turbulence~\cite{dunkel2013fluid} to flocking motion~\cite{vicsek2012collective}. These effects are revealed by the emergence of non-trivial correlations~\cite{cavagna2018physics}, either in the instantaneous velocity field or in displacement fields over larger time lags.

In systems of self-propelled particles without aligning interactions, sometimes known as scalar active matter, the difference between active and passive systems appears through the persistence time $\tau_p$ of the self-propulsion~\cite{berthier14,szamel2014self,fodor2016how}. The passive behaviour can be recovered in the non-persistent limit $\tau_p \to 0$.  The other extreme of very large values of $\tau_p$ leads to distinct behaviour, such as jamming and intermittent relaxation~\cite{mandal2020extreme,mandal2021how,keta2022disordered}, reminiscent of sheared athermal systems~\cite{maloney2006amorphous}. In such persistent systems, collective motion emerges when crowding effects at large density compete with persistent activity~\cite{szamel2015glassy,flenner2016nonequilibrium,berthier2017active,henkes2020dense,keta2022disordered}.  This happens in both fluid and glassy states, and increasing the persistence time typically reinforces these effects~\cite{mandal2020extreme,henkes2020dense,szamel2021long} because the correlation length of velocity correlations grows with $\tau_p$. However, computational studies of this regime are arduous because of the separation of timescales between $\tau_p$ and emergent correlation times on the one hand, and molecular timescales on the other.

For very large but finite $\tau_p$, it has recently been shown that this difficulty can be tackled using a strategy known as activity-driven dynamics (ADD)~\cite{mandal2021how,mandal2020multiple}. For systems without thermal noise, this provides direct access to the limit of very large $\tau_p$, and allows simulations of all dynamical relaxation processes taking place on timescales comparable with and greater than $\tau_p$. The central idea is that for times $t\ll \tau_p$, the system evolves as if the propulsion forces were fixed: if these forces are not too strong, then the dynamics will converge to a force-balanced configuration in a time $\tau_0 \ll \tau_p$.  
For larger forces, the system yields and there is no convergence to any mechanical equilibrium~\cite{liao2018criticality,mandal2020extreme,villarroel2021critical}; ADD is restricted to forces below that yield point.
After reaching a force-balanced state, the system configuration in ADD remains almost constant until the propulsion forces change significantly, which requires a time of order $\tau_p$.  Eventually, these forces will change sufficiently to destabilise the force-balanced state, at which point the system must switch to a new force-balanced configuration, which involves significant particle motion.

The resulting intermittent dynamical motion has several important consequences. First, it means that standard simulation methods become inefficient because they require a time step $\delta t \lesssim \tau_0 \ll \tau_p$.  ADD circumvents this problem by  replacing the explicit dynamical integration over times of order $\tau_0$ with an energy minimisation step.  This offers a computational speedup of order $\tau_p/\tau_0$, which is large in extremely persistent active matter.  Second, the intermittent mechanism of ADD relaxation -- which involves long quiescent periods punctuated by large sudden motions -- means that these systems share similarities with other physical systems where athermal quasistatic motion is relevant.  These include amorphous solids under athermal quasistatic shear (AQS)~\cite{maloney2006amorphous}, and active matter systems undergoing an athermal quasistatic random displacement (AQRD) protocol~\cite{morse2021direct}.

In all these contexts, a system evolves in response to potential gradients and the potential evolves quasistatically: the resulting motion is mostly quasistatic, but sudden relaxation events are triggered when the local minimum of the potential deforms into a saddle point.  Such events are known as avalanches, because large-scale motion can be triggered after an infinitesimal local change~\cite{sethna1993hysteresis,liu2016driving,ozawa2018random}.  

In addition to similarities with AQS and AQRD, the ADD method is also related to the athermal quasistatic random force (AQRF) protocol of Ref.~\cite{morse2021direct}.  However, ADD is distinct from all these methods. Specifically, both AQS and AQRD displace particles along a fixed driving direction and particles move orthogonally to the drive to minimise their interaction energy.  This leads to sustained stick-slip motion, including avalanches.  By contrast, AQRF applies forces with fixed direction; these are changed smoothly, which generates avalanches.  However, increasing the force amplitude eventually drives the system through a yielding threshold~\cite{liao2018criticality,mandal2020extreme,villarroel2021critical}, after which mechanical equilibrium is no longer reached and the system ``flows''. The essential features of ADD are that it controls the forces on particles (contrary to AQS and AQRD); and that the directions of these forces change randomly with time while their typical strength remains constant, contrary to AQRF. The result is a dynamical non-equilibrium steady state that exhibits stick-slip motion.  This corresponds to the large-$\tau_p$ limit of the steady states previously observed in active matter, which also show highly intermittent motion~\cite{mandal2020extreme,keta2022disordered}.

In this work, we develop an efficient computational approach to exploit the ADD method in order to characterise the dynamical relaxation of dense active matter over timescales $t \gg \tau_p$ in the extremely persistent limit of $\tau_p \gg \tau_0$.  
We work in two dimensions, which is the relevant case for many experimental active systems~\cite{angelini2011glasslike,kumar2014flocking,dauchot2016crystal}, and also helps visualisation of the complex physics of active systems. We address two broad sets of questions.

First, we show that ADD relaxation is built on two fundamental processes as in other athermal quasistatic methods such as AQS. These are smooth elastic deformations where the particles deform weakly near an energy minimum, in response to the slowly-varying self-propulsion forces; and sudden plastic rearrangements or avalanches in which a local minimum of the energy landscape becomes unstable, forcing the system towards a new one. We characterise these two processes in detail, including quantitative comparisons with AQS. 

Our second broad question is how particles move and relax in the steady states of ADD.  As one might expect for a dense disordered system with slow dynamics, one finds cooperative heterogeneous relaxation.  We analyse this motion using a range of techniques borrowed from passive glassy systems~\cite{berthier2011dynamical}, including distributions of particle displacements, correlation functions tailored for structural relaxation, and four-point susceptibilities.  

Our results generate significant insight into active matter at large $\tau_p$.  For the elastic and plastic events, several of our results show spatial correlations that extend across the entire system. For example, the mean size of a plastic event grows with system size $N$~\cite{mandal2021how}, and we also observe a significant $N$-dependence of the yielding threshold~\cite{mandal2020extreme}. This should be contrasted with thermal glassy systems where one expects the behaviour of a large bulk system to be well-approximated by dividing it into several non-interacting subsystems. This behaviour appears because we choose to take the limit $\tau_p\to\infty$ at fixed system size, which implies that the characteristic length scale for velocity correlations grows with the system size \cite{henkes2020dense}. We discuss the consequences of this choice when comparing ADD dynamics with direct simulation of active matter. 

For structural relaxation mechanisms in the steady state, we find behaviour with many similarities to passive glasses, including dynamical slowing down, stretched exponential relaxation, and significant dynamical heterogeneity, including dynamical facilitation effects~\cite{chandler2011}. On the other hand, the standard two-step relaxation mechanism of thermal glasses~\cite{berthierbiroli11} and weakly persistent active glasses~\cite{berthier2019glassy} is not observed. There is no notion of thermal motion within a cage and particles instead move slowly and smoothly in elastic trajectory segments before suddenly relaxing by plastic ones. The second major distinction with thermal systems is the existence of system-spanning dynamic correlations that control the behaviour of time correlations functions in a way that is again qualitatively reminiscent of AQS in sheared glasses.  

The remainder of the paper is organised as follows. Sec.~\ref{sec:model} defines the model and presents our implementation of the ADD method. Sec.~\ref{sec:single} characterises the elastic and plastic steps. Sec.~\ref{sec:dyn} analyses the structural relaxation and Sec.~\ref{sec:conc} collects our conclusions.

\section{Model and methods}

\label{sec:model}

\subsection{Model}

We consider a system of $N$ polydisperse overdamped active Ornstein-Uhlenbeck particles (AOUPs) in a two-dimensional periodic box of linear size $L$.  This is the same system as in~\cite{keta2022disordered} where large finite $\tau_p$'s were considered; here we use ADD to access the limit $\tau_p\to\infty$. Particle $i$ has position $\bfr_i$ and diameter $\sigma_i$; it feels a self-propulsion force $\bfp_i$ with persistence time $\tau_p$; and it interacts with other particles through a regularised soft-sphere Weeks-Chandler-Andersen potential of strength $\varepsilon$  (see SM).
The resulting equations of motion are
\begin{align}
\label{orig-aoupr}
\xi \dot{\boldsymbol{r}}_i(t) & = -\nabla_i U(t) + \boldsymbol{p}_i(t) , \\
\label{orig-aoupp}
\tau_p \dot{\boldsymbol{p}}_i(t) & = - \boldsymbol{p}_i(t) + \sqrt{2 \xi^2 D_0} \, \boldsymbol{\eta}_i(t) ,
\end{align}
where $\xi$ is a viscous damping coefficient, $D_0$ is the free-particle diffusion constant, and each component of the vector $\boldsymbol{\eta}_i$ an independent zero-mean unit-variance Gaussian white noise. The propulsions $\boldsymbol{p}_i$ follow a non-interacting Ornstein-Uhlenbeck process, and thus their distribution is Gaussian at each time. Particle diameters are drawn from a uniform distribution of mean $\sigma=\overline{\sigma}_i$ and polydispersity 20\% \cite{fily2014freezing,keta2022disordered}.
The motivation for studying polydisperse systems is to suppress the tendency of these systems to crystallise \cite{digregorio2018full}, and to allow access to the dense fluid regime where motion is complex and cooperative~\cite{keta2022disordered}.

At fixed propulsion forces, the system converges to a force-balanced configuration in a time $\approx \tau_0 = \xi \sigma^2 / \varepsilon$ if these forces are not too strong~\cite{mandal2021how}.
We focus here on extremely persistent systems where $\tau_p \gg \tau_0$.
In this limit, it is convenient to introduce a rescaled time variable as
\begin{equation}
t^{\prime} = t/\tau_p \; .
\end{equation}
We also rescale the propulsion by defining $\tilde{\bfp}=\bfp/f$ with
$f = \sqrt{\xi^2D_0/\tau_p}$.  It is convenient to work in the centre-of-mass frame so we write $\overline{\tilde{\bfp}} = (1/N) \sum_i \tilde{\bfp}_i$ and $\overline{\boldsymbol{r}} = (1/N) \sum_i \boldsymbol{r}_i$, which are the propulsive force acting on the particles' centre of mass and the position of the latter.
In the centre-of-mass frame and using the above reduced units, Eqs.~(\ref{orig-aoupr},\ref{orig-aoupp}) become
\begin{align}
\label{scaled-aoupr}
\frac{\xi}{\tau_p} \frac{\mathrm{d}\boldsymbol{r}^{\prime}_i}{\mathrm{d}t^{\prime}}(t^{\prime}) & =  -\nabla_i U(t^{\prime}) + f  [\tilde{\boldsymbol{p}}_i(t^{\prime}) - \overline{\tilde{\bfp}}(t^\prime)]  , \\
\label{scaled-aoupp}
\frac{\mathrm{d}\tilde{\boldsymbol{p}}_i}{\mathrm{d}t^{\prime}}(t^{\prime}) & = -\tilde{\boldsymbol{p}}_i(t^{\prime}) + \sqrt{2} \, \boldsymbol{\eta}^{\prime}_i(t^{\prime}) ,
\end{align} 
where $\boldsymbol{r}^{\prime}_i = \boldsymbol{r}_i - \overline{\boldsymbol{r}}$ is the position relative to the centre of mass, and $\boldsymbol{\eta}^{\prime}_i$ is a zero-mean Gaussian white noise in the rescaled time variables, that is, $\boldsymbol{\eta}^{\prime}_i = \sqrt{\tau_p} \boldsymbol{\eta}_i$, which ensures that the components of $\boldsymbol{\eta}^{\prime}_i$ satisfy $\langle \eta'_i(t_1')\eta'_i(t_2')\rangle = \delta(t_1'-t_2')$.

To arrive at the ADD limit, we take limit $\tau_p \to \infty$ at fixed $f$. For large $\tau_p$, Eq.~(\ref{scaled-aoupr}) describes very fast relaxation to configurations with perfect force balance, which satisfy
\begin{equation}
\nabla_i U_{\rm eff} = 0,  
\label{dUeff0}
\end{equation}
where 
\begin{equation}
U_{\rm eff} = U - f \sum_j [\tilde{\boldsymbol{p}}_j(t^{\prime}) - \overline{\tilde{\bfp}}(t^\prime)]\cdot\boldsymbol{r}_j
\label{Ueff-aoup}
\end{equation}
is an effective potential, which corresponds to the original potential energy of the interacting particle system, tilted by the active forces.  Hence, as $\tau_p\to\infty$, the system is almost always in a local minimum of $U_{\rm eff}$, as captured mathematically by Eq.~(\ref{dUeff0}).

It is natural to fix $\sigma$ as the unit of length and $\varepsilon$ as the unit of energy.  The rescaled time $t'$ is already dimensionless.  In the ADD limit $\tau_p\to\infty$ studied here, the only remaining control parameters are therefore the number density $\rho=N/L^2$, the self-propulsion force $f$, and the total number of particles $N$. We work throughout at the representative density $\rho = 1.2$. For fixed $f$, we expect the behaviour to be robust with respect to $\rho$, as long as $\rho$ is not too large (complete jamming) or too small (no force balanced states, i.e.\ flow).

\subsection{The quasi-static ADD method}

The dynamics of our model can be simulated directly in the ADD limit \cite{mandal2021how}; we summarise here the method for achieving this.
The steady state distribution of the propulsions $\tilde{\bfp}_i$ factorises across particles, with
\begin{equation}
P(\tilde{\boldsymbol{p}}_i) = \frac{1}{2 \pi} \exp\left(-\frac{1}{2} |\tilde{\boldsymbol{p}}_i|^2\right) ,
\end{equation}
and we use this to initialise the $\tilde{\boldsymbol{p}}_i$.
In each step of the ADD simulation, the
propulsion dynamics in Eq.~(\ref{scaled-aoupp}) is first integrated with time step $\delta t^{\prime}$:
\begin{equation}
\tilde{\boldsymbol{p}}_i(t^{\prime} + \delta t^{\prime}) = (1 - \delta t^{\prime}) \tilde{\boldsymbol{p}}_i(t^{\prime}) + \sqrt{2 \,\delta t^{\prime}} \, \tilde{\boldsymbol{\eta}}^{\prime}_i ,
\end{equation}
where $\tilde{\boldsymbol{\eta}}^{\prime}_i = (\tilde{\eta}^{\prime}_{i, x}, \tilde{\eta}^{\prime}_{i, y})$ are two random numbers drawn from a Gaussian distribution with zero mean and unit variance.

Next one integrates Eq.~(\ref{scaled-aoupr}), which requires that $U_{\rm eff}$ is minimised by steepest descent, holding the $\tilde{\bfp}_i$ fixed.
To increase computational efficiency, we replace this steepest descent by a faster conjugate gradient minimisation, using the GPL-licensed ALGLIB C++ library~\cite{alglib}. This is much more efficient than steepest descent, but it may generally lead to different local minima~\cite{nishikawa2022relaxation}. We find that the differences between conjugate gradient and steepest descent algorithms are significant only in steps for which the system moves far from its initial position. For these individual steps, we then automatically revert to steepest descent, as originally proposed in Ref.~\cite{mandal2021how}. That is, we choose a threshold in the mean-squared displacement $(1/N) \sum_i |\Delta \boldsymbol{r}_i|^2 > 0.1$ to identify minimisation steps with large total displacements. If this occurs during conjugate gradient minimisation then we restart the minimisation step and use steepest descent for that particular step. It is noteworthy that each of these minimisation step requires on average $10^2$-$10^3$ force evaluations. It is thus numerically challenging to explore both large systems and the large times needed to reach steady state. We will therefore show data for $N \leq 2000$.

The ADD construction is valid for forces $f$ below an $N$-dependent yielding threshold $f^*(N)$.  The potential $U_{\rm eff}$ is not bounded below so steepest descent may not converge to a local minimum -- instead the particles could continue to move along their self-propulsion directions. This happens for $f > f^*$; the rheology and phase behaviour of the system above this threshold is explored in Ref.~\cite{villarroel2021critical}. We locate the threshold $f^*(N)$~\cite{liao2018criticality} by letting the system evolve from a random initial configuration and checking for the average proportion of systems still flowing after some time $t$. We obtain the following rough estimates: $f^*(N=500) \approx 1.7$, $f^*(N=1024) \approx 1.4$, $f^*(N=2000) \approx 1.2$, $f^*(N=4096) \approx 1.1$.  
Given these systematic finite-size effects, it would be desirable to simulate even larger systems.  However, we emphasise that a single time step for ADD is much more expensive than a single time step in a standard molecular dynamics simulation, because it requires many evaluations of the interparticle forces to converge the energy minimisation. This stems from the intrinsic difficulty of simulating systems with well-separated time scales ($\tau_p/\tau_0\to\infty)$. This limit is inaccessible using standard methods: it can be simulated using ADD, but there is still a significant cost.

\begin{figure}
\includegraphics[width=\columnwidth]{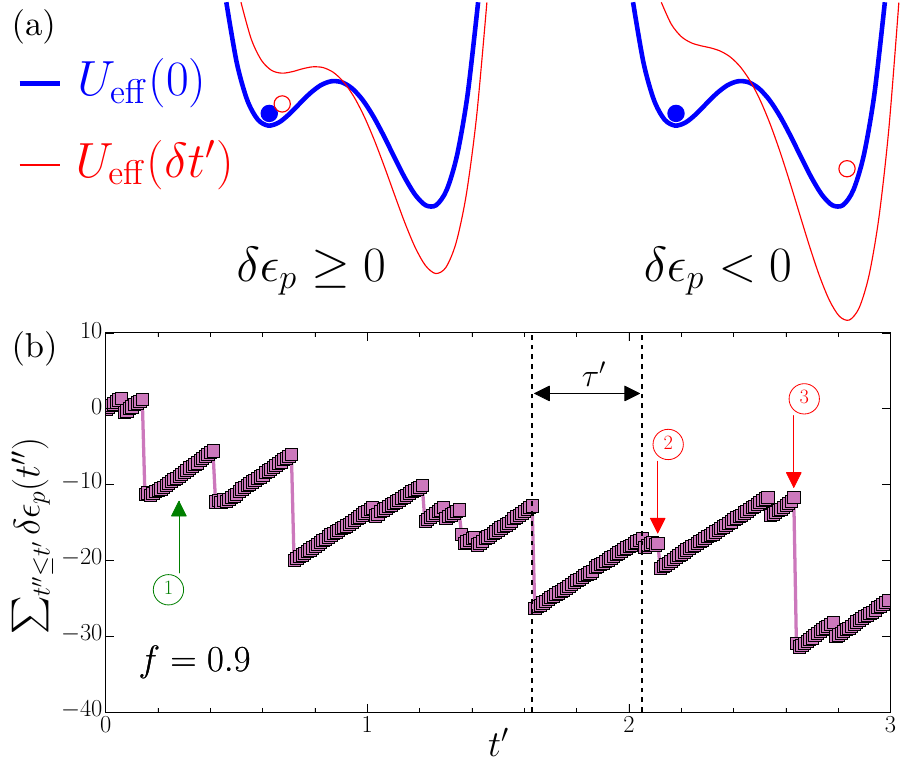}
\caption{(a) Sketch of the effective potential energy landscape $U_{\mathrm{eff}}$ in Eq.~(\ref{Ueff-aoup}) at times 0 (thick blue line) and $\delta t^{\prime}$ (thin red line). The system initially rests in a minimum of $U_{\mathrm{eff}}(0)$ (filled blue circle). After $\delta t^{\prime}$, the system rests in a minimum of the new landscape $U_{\mathrm{eff}}(\delta t^{\prime})$, and it is displayed in the initial landscape (open red circle). We distinguish elastic events ($\delta \epsilon_p \geq 0$) for which the systems remains close to its original position, and thus the potential energy in $U_{\mathrm{eff}}(0)$ increases, and plastic events ($\delta \epsilon_p < 0$) for which a rearrangement occurs.
(b) Accumulated variations of the effective potential energy, $\sum_{t''\leq t^{\prime}} \delta \epsilon_p(t'')$. We identify elastic branches made of successive elastic events (\textit{e.g.}\ event \Circled{1}) as ascending lines, and plastic events (\textit{e.g.}\ events \Circled{2} and \Circled{3}) as instantaneous large drops. We define $\tau^{\prime}$ as the time between two consecutive plastic events. Parameter values: $f=0.9$, $N=1024$, $\delta t^{\prime} = 10^{-2}$, purple square symbols separated by $\delta t^{\prime}$.}
\label{fig:dep}
\end{figure}

In the ADD dynamics for $f<f^*$, each step starts with the system in a local minimum of $U_{\rm eff}$ with positions and propulsions $(\bfr^0,\bfp^0)$, and evolves to a new minimum with positions and propulsions $(\bfr,\bfp)$. This can happen in two ways, as sketched in Fig.~\ref{fig:dep}(a).  In the simplest case, a small change in propulsive forces changes the local minimum of $U_{\rm eff}$ perturbatively, leading to small displacements. This will be called an elastic step.  However, the change in propulsive forces can also destabilise the local minimum at $(\bfr^0,\bfp^0)$, leading to a non-perturbative change in the configuration. This is called a plastic step. 

To distinguish these two cases, we compute the change in effective potential in one step:
\begin{equation}
\delta \epsilon_p = U(\boldsymbol{r}) - U(\boldsymbol{r}^0) - f \sum_i (\tilde{\boldsymbol{p}}_i^0 - \overline{\tilde{\boldsymbol{p}}^0}) \cdot (\boldsymbol{r}_i - \boldsymbol{r}_i^0) .
\label{eq:dep}
\end{equation}
This sign convention is opposite to that of Ref.~\cite{mandal2021how}. Note that the propulsions in this equation are those of the state before the ADD step, with the consequence that perturbative changes in the positions lead to positive $\delta \epsilon_p$ as the system moves away from the minimum of the associated $U_{\rm eff}$. Hence, we identify elastic steps as those with $\delta \epsilon_p\geq 0$ while those with $\delta \epsilon_p < 0$ are plastic, as illustrated in  Fig.~\ref{fig:dep}(a). We discuss these two types of step separately in detail below.

As well as the system parameters, a numerical simulation of ADD also requires a choice of time step $\delta t^\prime$.  As usual, this  should be small enough to mimic the limit $\delta t^\prime\to0$, but large enough to ensure computational efficiency. In practice, we chose $\delta t'$ depending on the state point: we tested several choices and monitored the mean squared displacement during plastic events. We chose a value small enough that this quantity depends at most weakly on $\delta t'$~\cite{mandal2021how}.

\section{Analysis of individual events}

\label{sec:single}

A typical trajectory from ADD is shown in Fig.~\ref{fig:dep}(b).  It consists of sequences of elastic steps [$\delta \epsilon_p$ is positive and $O(\delta t^{\prime})$], interspersed with instantaneous plastic events [$\delta \epsilon_p$ is negative and $O(1)$].  Such behaviour is familiar from AQS simulations of sheared glasses~\cite{maloney2006amorphous} as well as from AQRD simulations~\cite{morse2021direct} and the non-equilibrium dynamics of the random-field Ising model (RFIM, where the plastic events would be identified as avalanches)~\cite{ozawa2018random}.

This section analyses the properties of the plastic and elastic steps, including a comparison with AQS.  We take $f=0.9$ throughout this section.
This is a practical choice: on the one hand it is far enough from the threshold $f^*$ to keep the system from moving too much between minimisations; on the other hand smaller values of $f$ lead to slower dynamics and the numerics become more challenging, as discussed in Sec.~\ref{sec:dyn}.

\subsection{Elastic steps}

\begin{figure}
\includegraphics[width=\columnwidth]{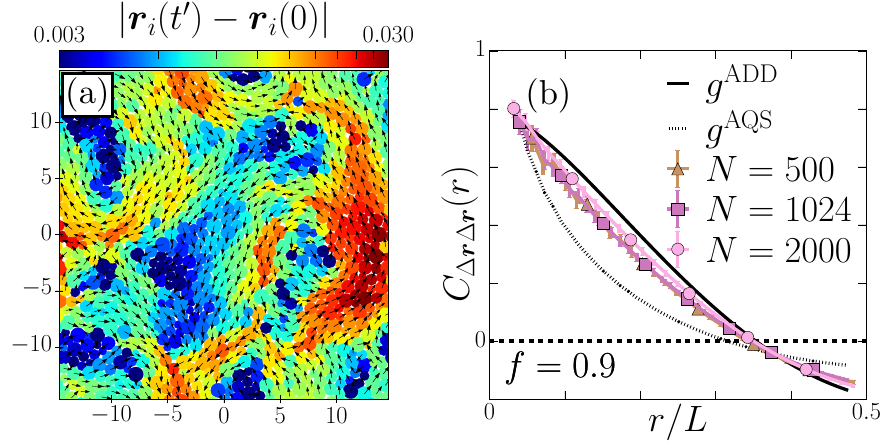}
\caption{(a) Snapshot of displacements for a single elastic step (\Circled{1} in Fig.~\ref{fig:dep}). Colours indicate the norm and arrows their direction.
Parameter values: $f=0.9$, $N=1024$, $\delta t^{\prime} = 10^{-2}$.
(b) Corresponding displacement correlation function $C(r) = \left< \delta \boldsymbol{r}_i  \cdot \delta\boldsymbol{r}_j \delta(r - r_{ij})\right>/\left<|\delta \boldsymbol{r}_i|^2 \right>$ for various $N$ values. The collapse with $r/L$ for three different system sizes shows that correlations scale with system size, becoming negative for large enough $r/L$.
Solid (resp.\ dotted) line: evaluation of \eqref{eq:gadd} (resp.\ \eqref{eq:gaqs}) with sums across the range $0 < m^2 + n^2 < 40^2$ and a suitably chosen overall prefactor.}
\label{fig:elastic}
\end{figure}

A representative snapshot of the displacement field obtained during an elementary elastic step of the ADD dynamics is shown in Fig.~\ref{fig:elastic}(a). We observe highly heterogeneous displacements, with wide variations in amplitude and clear large-scale correlations resembling both non-affine displacement in sheared athermal glasses and collective swirling motion in active matter \cite{henkes2020dense}.

Elastic steps can be analysed under the assumption that the updated propulsive forces move the minimum of $U_{\rm eff}$ perturbatively, in a way similar to AQS~\cite{maloney2006amorphous,maloney2006correlations}.
The states before and after the elastic step are both force balanced, so for all $i$
\beq
\delta 
\left[-\nabla_i U + f (\tilde{\boldsymbol{p}}_i-\overline{\tilde{\bfp}})\right] = 0 ,
\eeq
where $\delta$ indicates the change in a time increment $\delta t'$.
For small $\delta t'$ this implies that
\beq
-\sum_j (\mathbb{H})_{ij} \delta \boldsymbol{r}_j + \boldsymbol{\Xi}_i = 0 ,
\label{eq:harm}
\eeq
where $\mathbb{H}$ is the Hessian matrix of $U$, that is $(\mathbb{H})_{ij} = \frac{\partial^2 U}{\partial \bfr_i \partial \bfr_j}$, and 
\begin{equation}
\begin{aligned}
\boldsymbol{\Xi}_i & = f (\delta \tilde{\boldsymbol{p}}_i-\delta \overline{\tilde{\bfp}})
\\ &= f\, (-\tilde{\boldsymbol{p}}_i \delta t'+ \sqrt{2\,\delta t^{\prime}} \boldsymbol{\eta}_i^{\prime}-\delta \overline{\tilde{\bfp}})
\end{aligned}
\label{eq:dforce_affine}
\end{equation}
is the analogue of the affine force in the AQS setting of Ref.~\cite{maloney2006amorphous}, where the term $\delta \overline{\tilde{\bfp}}$ only ensures $\sum_i \boldsymbol{\Xi}_i=0$ so that we stay in the centre-of-mass frame. 

To evaluate the solution of \eqref{eq:harm} approximately, we  follow Ref.~\cite{maloney2006amorphous} in assuming that the eigenmodes of the Hessian $\mathbb{H}$ can be decomposed as plane wave eigenmodes of the Navier operator \cite{slaughter2002linearized} (see SM)
\begin{equation}
\begin{aligned}
\delta \boldsymbol{r}_i =& \sum_j (\mathbb{H}^{-1})_{ij} \boldsymbol{\Xi}_{j}\\
=& \sum_{m, n, \alpha} \frac{\Xi_{mn}^{\alpha}}{\lambda^{\alpha}(m^2 + n^2)} e^{\mathrm{i} \boldsymbol{k}_{m n} \cdot \boldsymbol{r}_i} \hat{\boldsymbol{k}}^{\alpha}_{m n} ,
\end{aligned}
\label{eq:drvec}
\end{equation}
where $\Xi_{mn}^{\alpha}$ is the projection of the affine force along the corresponding eigenmode, $\boldsymbol{k}_{m n} = (2\pi m/L, 2\pi n/L)$ is the wave vector and the third sum is over the two polarization directions of the elastic displacements $\alpha=||,\perp$ (longitudinal and transverse, respectively). The corresponding polarization vectors $\hat{\boldsymbol{k}}_{m n}^\alpha$ are the 
unit vectors parallel and orthogonal to the wave vectors, respectively. We have written the associated eigenvalues as $\lambda^\alpha(m^2+n^2)$, with the prefactor again depending on the polarization direction. It is then possible to compute the spatial correlation function
\begin{equation}
\begin{aligned}
g^{\mathrm{ADD}}(\boldsymbol{r}) \propto& \left<\delta \boldsymbol{r}_i \cdot \delta \boldsymbol{r}_j \, \delta(\boldsymbol{r} - (\boldsymbol{r}_j - \boldsymbol{r}_i))\right>\\
=& \sum_{m, n, \alpha} \frac{\left<|\Xi_{mn}^{\alpha}|^2\right>}{(\lambda^{\alpha} (m^2 + n^2))^2} e^{\mathrm{i} \boldsymbol{k}_{m n} \cdot \boldsymbol{r}} ,
\end{aligned}
\label{eq:gaddv}
\end{equation}
where we have checked numerically that the projections of the affine force $\boldsymbol{\Xi}_i$ on the eigenmodes of the Hessian $\mathbb{H}$ behave as uncorrelated random numbers, with a variance $\left<|\Xi_{mn}^{\alpha}|^2\right>$ independent of the specific mode.
We finally take the orientational 
%cylindrical 
average of \eqref{eq:gaddv} to write
\begin{equation}
g^{\mathrm{ADD}}(r) \propto \sum_{m, n} \frac{J_0\left(2\pi \sqrt{m^2 + n^2} \, r/L\right)}{(m^2 + n^2)^2} ,
\label{eq:gadd} 
\end{equation}
where $J_0$ is the zeroth-order Bessel function of the first kind. Our result is quantitatively different from the corresponding correlation function for sheared amorphous solids~\cite{maloney2006correlations}
\begin{equation}
g^{\mathrm{AQS}}(r) \propto \sum_{m, n} \frac{J_0\left(2\pi \sqrt{m^2 + n^2} \, r/L\right)}{m^2 + n^2} \; .
\label{eq:gaqs}
\end{equation}
The difference between (\ref{eq:gadd},\ref{eq:gaqs}) arises because affine forces $\boldsymbol{\Xi}_i$ in AQS are derived from pair potentials: the force exerted by particle $i$ on particle $j$ is equal and opposite to the one exerted by particle $j$ on particle $i$.  Hence, the affine forces on different particles are necessarily correlated in AQS. Still, both functions correspond to scale-free correlations, with the only relevant length scale being the system size $L$ itself. Qualitatively, this behaviour is visible already from the system-spanning vortices in the displacement field shown in Fig.~\ref{fig:elastic}(a).

We compute the displacement correlation quantitatively in Fig.~\ref{fig:elastic}(b).
These correlations scale with the system size for the range of sizes we have investigated, and are close to the analytical prediction \eqref{eq:gadd}.
The small difference between the prediction and the measurement may be attributed to the plane wave hypothesis, which can in principle be tested~\cite{mizuno2017continuum}. We have also checked that the above correlation functions are consistent with the correlations computed from purely harmonic steps, i.e.\ with displacements determined by solving \eqref{eq:harm} exactly. This establishes another parallel to Ref.~\cite{maloney2006correlations}.

The scaling with system size of the displacement correlations that we find is consistent with the arguments obtained for finite persistence time in Ref.~\cite{henkes2020dense}. In that case, the dynamics along elastic trajectory segments produces displacement (or equivalently velocity) correlations on a length scale that diverges as $\sim\sqrt{\tau_p}$ for large persistence times. As our analysis takes $\tau_p \to \infty$ from the start, this limit translates into displacement correlations on the largest length scale available, {\em i.e.}\ the system size.

\subsection{Plastic steps}

\begin{figure}
\includegraphics[width=\columnwidth]{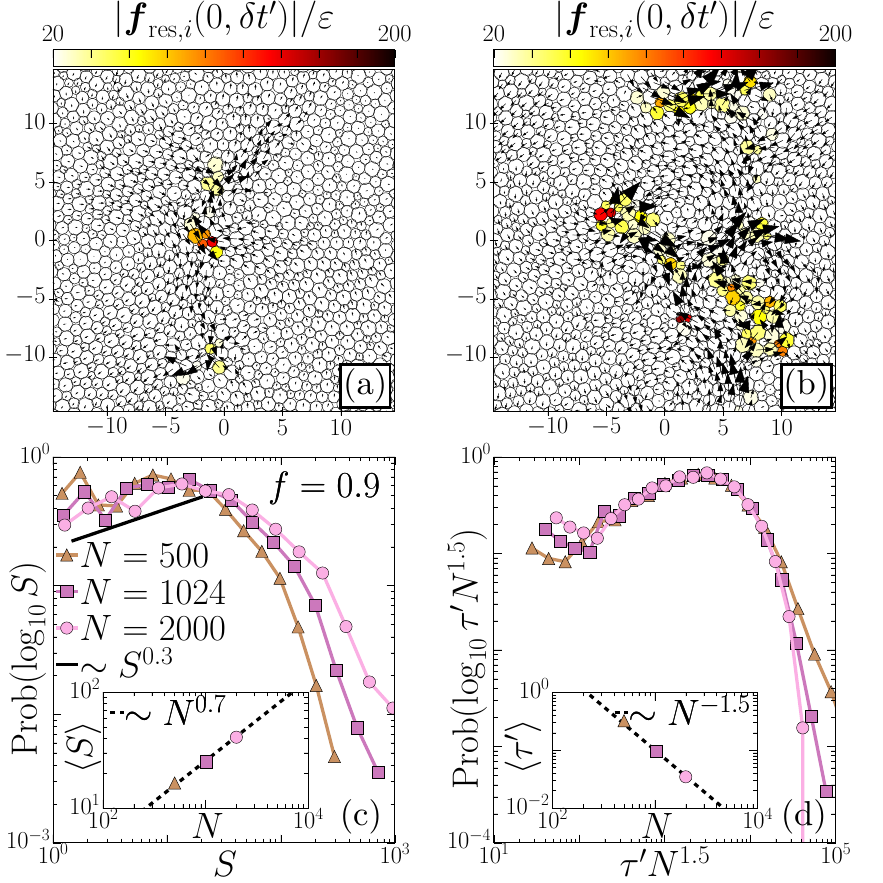}
\caption{(a,b) Snapshots showing the movement of particles in the two plastic events marked \Circled{2} and \Circled{3} in Fig.~\ref{fig:dep}, with participation $S=19$ and $S=102$, respectively. Displacements are magnified 5 times and superimposed onto a colormap of the residual force $|\boldsymbol{f}_{\mathrm{res},i}|$ that highlights rearranging particles. Parameter values: $f=0.9$, $N = 1024$, $\delta t^{\prime} = 10^{-2}$.
(c) Log-distribution of the participation $S$ in plastic events, for three system sizes $N$ and time steps $\delta t^{\prime}(N=500) = 2 \times 10^{-3}$, $\delta t^{\prime}(N=1024) = 10^{-3}$, $\delta t^{\prime}(N=2000) = 5 \times 10^{-4}$.
The solid black line corresponds to a scaling $P(S) \sim P(\log_{10} S)/S \sim S^{0.3 - 1} = S^{-0.7}$.
Inset shows the evolution of the mean $\langle S \rangle$ with $N$.
(d) Log-distribution of times $\tau^{\prime}$ between consecutive plastic events. Inset shows the evolution of the mean $\langle \tau' \rangle$ with $N$.
Scaling exponents are subject to significant uncertainties and the numbers provided are indicative.}
\label{fig:plastic}
\end{figure}

Particle displacements for two representative plastic steps are shown in Figs.~\ref{fig:plastic}(a,b). The qualitative picture has again many similarities with AQS: the displacements in any single plastic event can be interpreted as sequences of localised yielding events~\cite{mandal2021how,maloney2006amorphous}.  That is, a plastic step happens when a local minimum of $U_{\rm eff}$ develops an unstable direction, causing local motion; but the elastic perturbation due to this event perturbs the system over large length scales and can create further unstable directions in other parts of the system.  This leads to a cascade or avalanche of localised yielding events. We describe these plastic events here and discuss similarities and differences with AQS.

We find that plastic avalanches display a broad range of sizes and can involve a few localised particles, as in Fig.~\ref{fig:plastic}(a), or a greater number of particles distributed across the system as in Fig.~\ref{fig:plastic}(b), or even the whole system. To characterise the participation in each plastic event, we identify the %particles  participate in the events, and denote 
number $S$ of particles with significant changes in their local environments. In Ref.~\cite{mandal2021how} this identification was carried out by thresholding particle displacements, but such a criterion neglects the fact that particles may collectively move large distances without changing their local environment.  Here, we use instead the residual force $\boldsymbol{f}_{\mathrm{res},i}$~\cite{lemaitre2014structural,lerbinger2021relevance} (see SM for definition) as an indicator of rearrangements.
This force is zero if and only if displacements result from a harmonic response of the system to the change in propulsion forces.  If a localised avalanche takes place, particles that are far from the avalanche tend to respond elastically, leading to very low residual forces, whereas rearranging particles in the core of the avalanche have large residual forces.  This is illustrated in Fig.~\ref{fig:plastic}(a,b). We then define the avalanche size $S$ as the number of particles for which $|\boldsymbol{f}_{\mathrm{res},i}(\delta t^{\prime})| > 20$, determined after careful analysis of the distributions of $\boldsymbol{f}_{\mathrm{res},i}$ (see SM).

Fig.~\ref{fig:plastic}(c) shows the resulting broad distribution of $\log_{10} S$ for three system sizes. The small events can be attributed to local yielding, in which the remainder of the system reacts elastically with particles moving collectively to accommodate the local rearrangement~\cite{mandal2021how,lerbinger2021relevance}. At low values of $S$, the distributions overlap for different system sizes, with a behaviour compatible with $P(S) \sim S^{-\tau}$ with $\tau \approx 0.7$. (We use the notation $\tau$ here, which is standard in the literature \cite{karimi2017inertia}; $\tau$ does not indicate a time scale.) Given the small range of system sizes studied here, it is difficult to provide a very precise estimate of $\tau$, but it is clearly distinct from the values found in AQS simulations of sheared glasses where values in the range $\tau \approx 1.2-1.5$ have been reported \cite{liu2016driving,karimi2017inertia}.

Turning to the behaviour at large $S$ we observe that larger avalanches with $S\sim N$ are more frequent for larger systems, suggesting that these are also important in ADD. Correspondingly, the inset in Fig.~\ref{fig:plastic}(c) shows that the average event size $\langle S \rangle$ scales as $N^\gamma$ with $\gamma \approx 0.7$, showing that the mean avalanche size is indeed controlled by large avalanches that are limited by the system size only. In other words, the avalanches observed during plastic events also lead to scale-free dynamic relaxation events. 

Fig.~\ref{fig:plastic}(d) shows the distribution of waiting times $\tau^{\prime}$ between consecutive events.
The average time decreases with system size as $\langle \tau^{\prime} \rangle \sim N^{-1.5}$.  If localised yielding events happened independently in different parts of a large system, one would have a more trivial dependence on the system size, $\langle \tau^{\prime} \rangle \sim N^{-1}$: together with the $N$-dependence of $\langle S\rangle$,  this is another indication of long-ranged correlations, on the scale of the system size.  

Such scaling behaviour hints at critical phenomena. The force threshold $f^*(N)$ for yielding decreases with $N$ in our simulations, which is presumably also due to long-range correlations~\cite{liao2018criticality,mandal2020extreme}.  Since we increase $N$ at fixed $f$ while staying always below $f^*(N)$, some of the dependence on $N$ may arise because the larger systems are closer to yielding. Indeed, larger systems support larger events, which tend to relax the system more quickly; compare Fig.~\ref{fig:dynN}(b) below.

While the ADD plastic events share similarities with those of AQS, there are also some important differences to emphasise. In particular, in ADD there is no preferred direction and the system is isotropic (apart from a possible influence of the periodic boundary conditions, which we expect to be very weak). In AQS, on the other hand, rotational symmetry is broken because the system is always sheared in the same direction.
As a result, localised plastic events eventually organise into a line of slip, which leads to a subextensive scaling of event sizes $\left<S\right> \sim L \sim \sqrt{N}$ in the steady state~\cite{maloney2004subextensive,maloney2006amorphous}. These correlations also cause a reduction in the frequency of plastic events: the typical time (accumulated strain) between consecutive events scales as $\left<\tau^{\prime}\right> \sim 1/L \sim 1/\sqrt{N}$ (we recall it would be $\sim 1/N$ for independent events~\cite{maloney2006amorphous}). 

In short, ADD in its elastic steps produces displacements that are correlated on the scale of the system size $L$, in agreement with predictions for finite $\tau_p$~\cite{henkes2020dense}.  However, the master curve for displacement correlations against $r/L$ is different from the AQS case~\cite{maloney2006correlations} because the local ``affine'' forces $\boldsymbol{\Xi}_i$ lack the correlations that are present for AQS~\cite{maloneyprivate}. For plastic events, the isotropy of ADD also leads to larger event sizes $S$ and shorter inter-event times. The next question to be addressed is how the individual particles move in the active fluid, when observed over multiple time steps. 

\section{Microscopic dynamics}

\label{sec:dyn}

Since ADD is a computational tool to explore particle motion in dense active fluids, it is natural to study dynamical relaxation in ADD trajectories. To this end, we use observables developed for the analysis of relaxation in dense glassy systems~\cite{berthier2011dynamical}.  Such measurements have also been used to describe particle motion in sheared and active glasses, which all display heterogeneous and cooperative dynamics.   

\subsection{Mean squared displacement}

\begin{figure}
\includegraphics[width=\columnwidth]{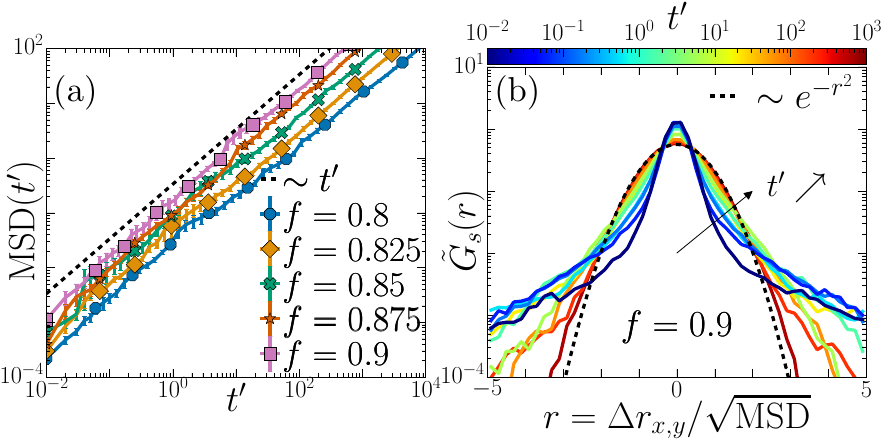}
\caption{(a) Mean-squared displacement for different values of the self-propulsion force $f$.
(b) Distribution of displacements scaled by the MSD at different $t'$. Parameter values: $N = 1024$, $\delta t^{\prime} = 10^{-2}$, $f=0.9$.}
\label{fig:dynf}
\end{figure}

Fig.~\ref{fig:dynf}(a) shows the mean squared displacement (MSD)
\begin{equation}
\mathrm{MSD}(t^{\prime}) = \left<|\boldsymbol{r}_i(t^{\prime}) - \boldsymbol{r}_i(0)|^2\right>
\label{eq:msd}
\end{equation}
for different values of the self-propulsion force $f$. In the steady state, the MSD is nearly diffusive at all times. The self-diffusion constant at large times roughly decreases by an order of magnitude between $f=0.9$ and $f=0.8$. However, there is no feature in the average displacements that would allow identification of a characteristic relaxation time scale or length scale. This is in contrast to the classic two-step relaxation scenario found in many glassy systems~\cite{keta2022disordered,binder2005glassy,berthierbiroli11}, but resembles the diffusive behaviour found in AQS simulations of sheared systems~\cite{heussinger2010superdiffusive}. 

\begin{figure}
\centering
\includegraphics[width=\columnwidth]{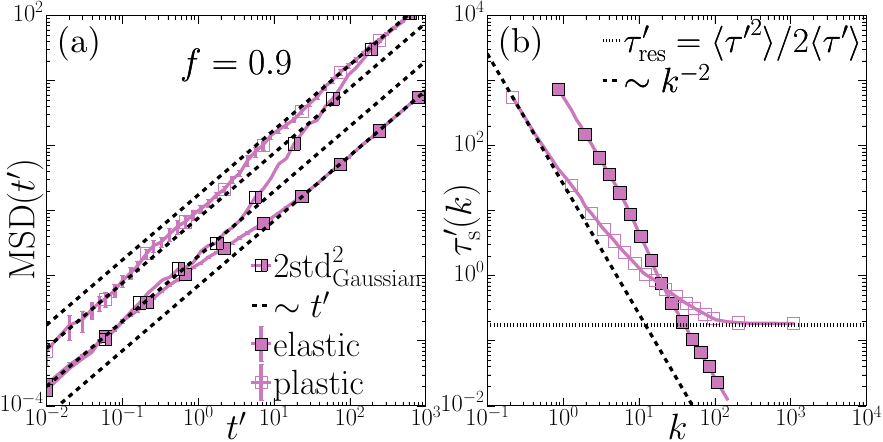}
\caption{(a) Elastic and plastic contributions to the MSD, together with the MSD predicted from a Gaussian fit of the central part of $G_s$. (b) Wave-vector-dependent relaxation time $\tau^{\prime}_{\mathrm{s}}(k)$ extracted from elastic and plastic displacements (colours and symbols as in (a)). Parameter values: $f=0.9$, $N = 1024$, $\delta t^{\prime} = 10^{-2}$.}
\label{fig:decomp}
\end{figure}

Although the MSD displays seemingly trivial behaviour, the displacement distributions have significant structure. Fig.~\ref{fig:dynf}(b) shows the corresponding distribution $G_s(r)$ of the $x$ and $y$-components of the particle displacements, scaled by the root mean-squared displacement, at $f = 0.9$. These distributions differ strongly from Gaussian behaviour, which is only recovered in the large time limit, $t' \to \infty$. At small times  $t^{\prime} \ll 1$ (with of course $t^\prime \geq \delta t^{\prime}$), the displacement distribution has a narrow central peak with heavy tails. The width of these tails decreases with increasing time, and the distribution approaches a Gaussian form. For supercooled liquids, we would expect the small time distribution to be nearly Gaussian due to short-time thermal dynamics, with fat tails developing only as the system starts to relax. The tails appear when a significant number of particle rearrangements has taken place~\cite{kob1997dynamical,chaudhuri2007universal}. The difference between ADD and thermal dynamics at short times is easily explained by the athermal quasistatic nature of ADD dynamics. Moreover, the participation in plastic events has a broad distribution [Fig.~\ref{fig:plastic}(c)]. As a result, the fat tails arising from structural relaxation are visible already after a single time step $\delta t'$ and arise from avalanches. In AQS simulations, similar heavy-tailed distributions also appear at early times due to plastic avalanches~\cite{heussinger2010superdiffusive}. In both ADD and AQS we expect that nearly-exponential tails appear at intermediate times, as a generic result of the stochastic nature of avalanches~\cite{chaudhuri2007universal}.

To gain more insight into the dynamics, we decompose the displacements into separate contributions from elastic and plastic events. We define the elastic (resp.\ plastic) displacement of a particle between $0$ and $t^{\prime}$ as the sum of its displacements over all elastic (resp.\ plastic) steps between these two times. We plot in Fig.~\ref{fig:decomp}(a) the MSDs from these contributions at $f = 0.9$. Both of them show a crossover between two diffusive scaling regimes, {\em i.e.}\ both have ${\rm MSD}(t')\sim t'$ at short and at long times but with different prefactors. Despite the complex time dependences of the separate contributions, their sum in the total MSD appears nearly linear (recall Fig.~\ref{fig:dynf}). 

To connect the elastic and plastic displacements to the distribution  ${G}_s(r)$ in Fig.~\ref{fig:dynf}(b), we fit the central peak of ${G}_s(r)$ to a Gaussian distribution with standard deviation $\mathrm{std}(t')$ such that the associated mean-squared displacement is $\mathrm{MSD}(t') = 2 \, \mathrm{std}^2(t')$. Fig.~\ref{fig:decomp}(a) compares this effective MSD to the elastic and plastic contributions. At small times, the central peak of ${G}_s$ is compatible with the variance of elastic displacements, while at large times it is compatible with plastic displacements. Our interpretation is that for small times, displacements in elastic branches populate the narrow central part of the displacement distribution, while the displacements of rearranging particles populate the tails.
At large times, plastic displacements dominate elastic displacements, and the whole displacement distribution is close to Gaussian, therefore its variance is dictated by plastic displacements.

\subsection{Single-particle correlation functions}

The MSD and the distribution $G_s$ yield useful information about the dynamics.  However, MSD measurements can sometimes be
 dominated by a subset of fast moving particles.  To investigate this, we computed the self-intermediate scattering function~\cite{binder2005glassy}
\begin{equation}
F_s(k, t^{\prime}) = \big<\cos\big[\boldsymbol{k} \cdot (\boldsymbol{r}_i(t^{\prime}) - \boldsymbol{r}_i(0))\big]\big>
\end{equation}
where $\boldsymbol{k}$ is a suitable wavevector and $k=|\boldsymbol{k}|$.  We analyse the $k$-dependent relaxation time scale $\tau^{\prime}_s(k)$~\cite{berthier2005length,berthier2004time}, which is defined as the time at which $F_s(k, \tau^{\prime}_s(k))=1/{\rm e}$.

We plot $\tau^{\prime}_s(k)$ for both elastic and plastic displacements in Fig.~\ref{fig:decomp}(b). Fickian diffusion would correspond to the scaling $\tau^{\prime}_s(k) \sim k^{-2}$.
Elastic displacements show two distinct diffusive scalings, at small and at large length scales (resp.\ small and large time scales).
At small times, elastic displacements are computed over a single elastic trajectory segment, thus the corresponding diffusive behaviour derives from the balance between the Ornstein-Uhlenbeck driving on the one hand and the restoring forces on the other hand \cite{henkes2012extracting,bottinelli2017how}. At large times, elastic displacements are computed over many elastic trajectory segments separated by multiple plastic events. We expect the displacements over these different elastic trajectory segments to be independent, therefore the sum of all these displacements produces a diffusive behaviour distinct from that of single branch displacements.

The small and large length scale behaviour are also different for plastic displacements. These displacements are indeed diffusive at large length scales but they show a relaxation time scale $\tau^{\prime}_s(k)$ that plateaus at small length scales. This plateau corresponds to the typical time for a plastic event to occur, and can be computed from the statistics of the inter-event times $\tau'$ as the residual time $\tau^{\prime}_{\mathrm{res}} = \langle {\tau^{\prime}}^2 \rangle / (2 \langle \tau^{\prime} \rangle)$ \cite{pal2022inspection}.
Therefore, at times $t^{\prime} \lesssim \tau^{\prime}_{\mathrm{res}}$ the plastic MSD is likely dominated by a small subset of particles.

It is noteworthy that this result is robust to changes in $f$ (data not shown). The residual time $\tau^{\prime}_{\mathrm{res}}$ changes by a factor of $\sim2.5$ between $f=0.9$ and $f=0.8$, distinct from the factor of $10$ observed for the self-diffusion constant (not shown). Moreover, since the diffusion constant drops more rapidly than the residual time increases, the typical length scale above which the plastic movement appears Fickian~\cite{berthier2005length} decreases with decreasing $f$ -- which is opposite to what we would expect for a supercooled liquid approaching the glass transition.

\begin{figure}
\includegraphics[width=\columnwidth]{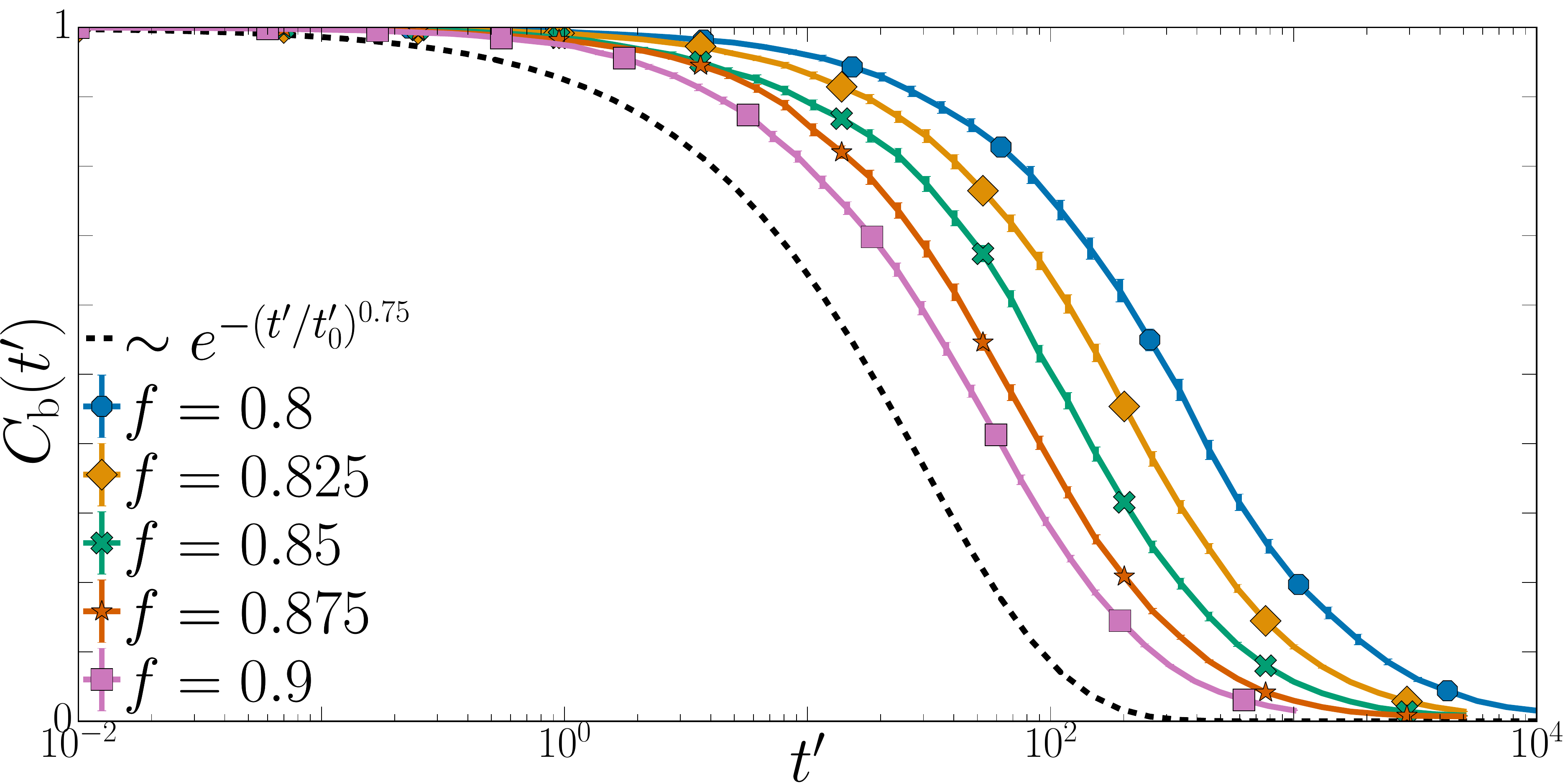}
\caption{Bond breaking correlation function $C_{\mathrm{b}}(t^{\prime})$ for different values of the self-propulsion force $f$.
Parameter values: $N = 1024$, $\delta t^{\prime} = 10^{-2}$.}
\label{fig:cbf}
\end{figure}

The displacement fields in the plastic events of Figs.~\ref{fig:plastic}(a,b) show that particles that are not involved in rearrangements can move away from their initial position without relaxing their local structure. But neither the MSD nor $F_s(k,t')$ can detect whether single particle translations actually correspond to changes in the local structure or not. To focus on this aspect of structural relaxation, we use the bond breaking correlation function $C_{\mathrm{b}}(t)$~\cite{shiba2012relationship}. Denoting $\hat{r}_{ij}(t^{\prime}) = |\boldsymbol{r}_j(t^{\prime}) - \boldsymbol{r}_i(t^{\prime})|/\sigma_{ij}$ as the rescaled distance between particles $i$ and $j$ at time $t^\prime$, with $\sigma_{ij}= (\sigma_i+\sigma_j)/2$, we define
\begin{equation}
C_{\mathrm{b}}(t^{\prime}) = \frac{\sum_{i,j} \Theta(A_1 - \hat{r}_{ij}(0)) \, \Theta(A_2 - \hat{r}_{ij}(t^{\prime}))}{\sum_{i,j} \Theta(A_1 - \hat{r}_{ij}(0))} \; ,
\label{eq:cb}
\end{equation}
where the parameter $A_1 = 1.25$ is a cutoff defining initial neighbours, $A_2 = 1.5$ quantifies the distance they are required to separate before the correlation function decays, and $\Theta$ designates the Heaviside function. This function obeys $C_b(t'=0)=1$, by definition, and it quantifies at time $t'$ the average fraction of neighbours lost since $t'=0$. This way, it efficiently disentangles rearrangements from displacements that do not relax the local structure.

Fig.~\ref{fig:cbf} shows $C_{\mathrm{b}}(t')$ for several values of $f$. The relaxation time scale $\tau_b^{\prime}$ of $C_{\mathrm{b}}$ roughly increases by an order of magnitude between $f=0.9$ and $f=0.8$, mirroring the decrease of the self-diffusion constant. Moreover, $\tau_b^\prime \gg 1$, so structural relaxation happens long after self-propulsion forces have fully decorrelated from their initial values which occurs for $t' \sim 1$. The correlation function is stretched, which suggests that structural relaxation is temporally heterogeneous~\cite{binder2005glassy}. In addition, it is remarkable that the MSD at the time $\tau_b^\prime$ where local structure becomes fully decorrelated is greater than unity. This is again very different from thermal glasses where the escape from the cage also coincides with structural relaxation. Here instead particles travel comparatively larger distance without necessarily relaxing the structure, which can be seen as a consequence of the swirling motion observed in snapshots such as Fig.~\ref{fig:elastic}(a) (for elastic events) or Fig.~\ref{fig:plastic}(a,b) (for plastic ones). 

\begin{figure}
\centering
\includegraphics[width=\columnwidth]{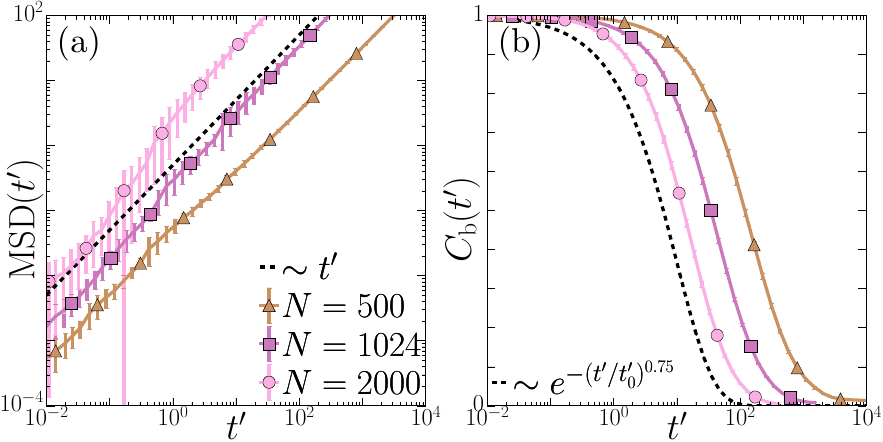}
\caption{(a) Mean-squared displacement and (b) bond-breaking correlation function for different $N$. Parameter values: $f=0.9$, $\delta t^{\prime}(N=500) = 2 \times 10^{-3}$, $\delta t^{\prime}(N=1024) = 1 \times 10^{-3}$, $\delta t^{\prime}(N=2000) = 5 \times 10^{-4}$.}
\label{fig:dynN}
\end{figure}

We finally show the dependence of the dynamics on system size $N$. Fig.~\ref{fig:dynN} shows the MSD and the bond breaking correlation function for a fixed $f$ but different values of $N$. For all values of $N$ studied, the MSD is diffusive at large times and the relaxation of $C_{\mathrm{b}}(t')$ is stretched. Strikingly, however, both functions strongly depend on the system size with no sign of a saturation at some large $N$ value. As $N$ increases, the particles move faster, resulting in an MSD that is larger and a time correlation function that decreases faster. Since we have simulated three system sizes, it is not easy to determine a precise scaling of the self-diffusion constant and the relaxation time $\tau'_b$ with $N$. There are two sources for the system size dependence of the dynamics as discussed above: the frequency and size of the plastic events both increase with $N$, recall Fig.~\ref{fig:plastic}. Thus, in the ADD regime, the dynamics is always sensitive to the system size, as a result of the large persistence time limit. This is again in good analogy with the AQS dynamics where the self-diffusion constant also changes with system size~\cite{lemaitre2007plastic}.

\begin{figure*}
\centering
\includegraphics[width=\textwidth]{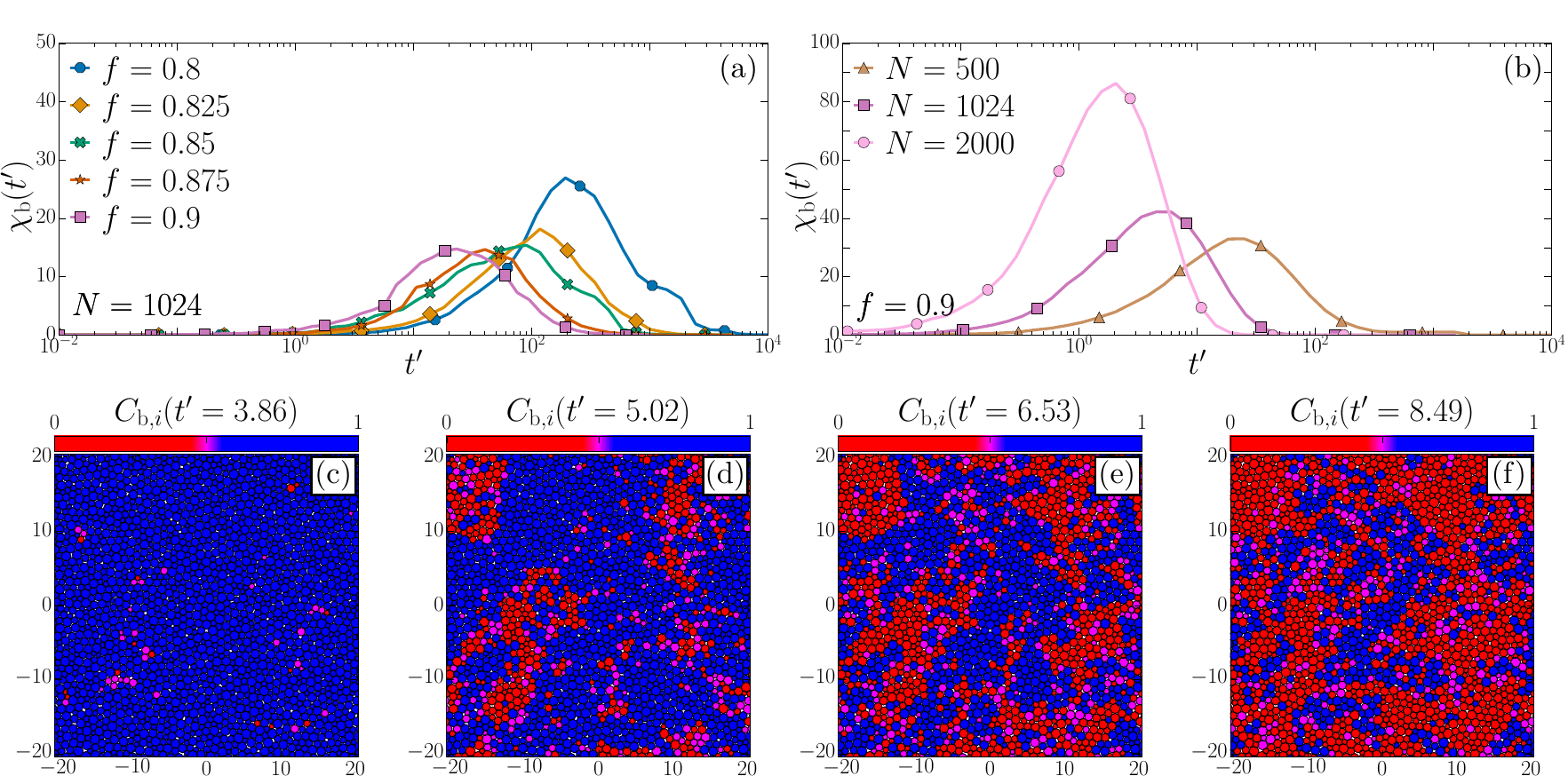}
\caption{(a) Bond breaking dynamical susceptibility $\chi_b(t^{\prime})$ for different self-propulsion forces $f$ with $N = 1024$ and $\delta t^{\prime} = 10^{-2}$. (b) Bond breaking dynamical susceptibility $\chi_b(t^{\prime})$ for different $N$ with $\delta t^{\prime}(N=500) = 2 \times 10^{-3}$, $\delta t^{\prime}(N=1024) = 1 \times 10^{-3}$, $\delta t^{\prime}(N=2000) = 5 \times 10^{-4}$. (c-f) Snapshots of the system highlighting the local bond breaking correlation $C_{\mathrm{b},i}$ between time $t'=0$ and a lag time of (c) $t^{\prime} = 3.86$ (124 plastic events, $C_{\mathrm{b}} = 0.89$) (d) $t^{\prime} = 5.02$ (196 plastic events, $C_{\mathrm{b}} = 0.65$) (e) $t^{\prime} = 6.53$ (252 plastic events, $C_{\mathrm{b}} = 0.52$) (f) $t^{\prime} = 8.49$ (355 plastic events, $C_{\mathrm{b}} = 0.41$). Parameter values: $N = 2000$, $f = 0.9$, $\delta t^{\prime} = 5 \times 10^{-4}$.}
\label{fig:heter}
\end{figure*}

\subsection{Dynamical heterogeneity}
\label{sec:heter}

In glassy fluids, one generally expects complex heterogeneous dynamics, where
spatial fluctuations around the average dynamical behaviour are important for understanding the relaxation dynamics \cite{berthier2011dynamical}. The system considered here also has this feature.  It is illustrated in the snapshots of Figs.~\ref{fig:heter}(c-f) which show maps of a single-particle analogue of the bond-breaking correlation function \eqref{eq:cb}. This is defined as
\beq
C_{\mathrm{b},i}(t^{\prime}) = \frac{\sum_j \Theta(A_1 - \hat{r}_{ij}(0)) \, \Theta(A_2 - \hat{r}_{ij}(t^{\prime}))}{\sum_j \Theta(A_1 - \hat{r}_{ij}(0))},
\label{eq:cbi}
\eeq
and represents the fraction of bonds of particle $i$ that have been broken up to time $t'$. Just like its global analogue, this function decays from unity to zero as the environment of particle $i$ decorrelates from its initial state.

The definition of a local relaxation function in Eq.~(\ref{eq:cbi}) allows us to visualise in real space how the initial structure of the system relaxes as dynamics proceeds. The main observation in the snapshots of Fig.~\ref{fig:heter} is that the spatial distribution of relaxed particles, at any given time, reveals strong spatial correlations in the local dynamics.

While similar observations of spatially correlated dynamics are quite generic in dense amorphous materials, the time series shown in Fig.~\ref{fig:heter}(c-f) reveals additional features beyond the mere existence of correlations. We observe that at short times [Fig.~\ref{fig:heter}(c)], only a few particles have relaxed, and the spatial structure of the $C_{b,i}$ reveals the very few underlying plastic events that have taken place in this particular trajectory. At larger times [Figs.~\ref{fig:heter}(d,e)], one sees that additional structural relaxation events tend to happen in the close vicinity of previous ones. These observations have also been made in slowly relaxing supercooled liquids at equilibrium and this effect is known as dynamical facilitation~\cite{guiselin2022microscopic,chandler2011}. At the microscopic level, these effects must correspond to spatio-temporal correlations between successive plastic events. Close to the relaxation time [Fig.~\ref{fig:heter}(e-f)], we can clearly identify fast regions where particles have been involved in numerous rearrangements and have relaxed their local initial structure.  We also see slow regions where particles' local environments remain the same. Finally, at long times all particles have $C_{b,i}\approx 0$ and one recovers a homogeneous picture. It is noteworthy that despite the absence of shear to organise plastic activity into anistropic structures (namely, shear bands~\cite{maloney2006amorphous}), correlations spontaneously emerge between the plastic centres. 

The extent of these dynamical spatial correlations can be quantified via the dynamical susceptibility \cite{shiba2012relationship,berthier2011dynamical,lacevic2003spatially}
\begin{equation}
\chi_{\mathrm{b}}(t^{\prime}) = N [\left<C_b(t^{\prime})^2\right> - \left<C_b(t^{\prime})\right>^2] \; .
\end{equation}
This function is plotted for different values of the self-propulsion force $f$ in Fig.~\ref{fig:heter}(a), and for different numbers of particles $N$ in Fig.~\ref{fig:heter}(b).
As usual, $\chi_{\mathrm{b}}$ shows a peak at the time close (but not exactly equal) to ${\tau_{\mathrm{b}}^{\prime}}$ where $C_b$ starts to decrease~\cite{shiba2012relationship}, indicating that the dynamics is most heterogeneous around these times. The slowdown of the dynamics with decreasing $f$ or $N$ (Figs.~\ref{fig:cbf},~\ref{fig:dynN}) is reflected in the corresponding increase of this peak time. Moreover, the height of the peak tells us about the typical number of particles involved in correlated clusters in Fig.~\ref{fig:heter} \cite{berthier2011dynamical}. At fixed $N$, the dynamical slowdown is accompanied by an increased cooperativity of the relaxation, as is observed for liquids approaching the glass transition~\cite{lacevic2003spatially}. This situation is different when the self-propulsion force $f$ is kept fixed: the dynamics speeds up with increasing $N$ but it also becomes more cooperative with a larger dynamical susceptibility. Recall that ADD is defined by taking the limit $\tau_p \to \infty$ at fixed $N$: as
a consequence, the length scale that characterises velocity fluctuations is slaved to the system size, and diverges for large $N$. The global correlation function $C_{\mathrm b}$ and its fluctuations $\chi_{\mathrm b}$ both change systematically with $N$, revealing that the long-time relaxation dynamics is also sensitive to the system size,   presumably because of a cooperativity length scale that diverges with $N$. 

These observations provide further evidence that spatially heterogeneous dynamics is very generic in dense and disordered fluids. A major difference with equilibrium supercooled liquids is the system size dependence observed for the dynamical heterogeneity, indicating a diverging correlation length.  This is attributed to the quasi-static nature of the dynamics, as also found in AQS simulations~\cite{heussinger2010superdiffusive}. In addition, the slow growth of dynamic correlations with time (Fig.~\ref{fig:heter}) reveals the role of dynamic facilitation. Whereas facilitation has been described before in equilibrium dynamics~\cite{guiselin2022microscopic,scalliet2022thirty,PhysRevX.1.021013}, much less is understood about its consequences for sheared and active systems. Our findings suggest that facilitation could also be a very generic feature characterising the relaxation dynamics of dense and disordered fluids, and this clearly deserves further study in the context of driven amorphous materials.

\section{Conclusion}

\label{sec:conc}

Although models of spherical self-propelled particles are among the simplest models of active matter they exhibit rich physics with emerging structures, phases and dynamical behaviours. A remarkable feature is the emergence of non-trivial velocity correlations in fluid~\cite{szamel2021long}, glassy~\cite{szamel2015glassy,henkes2020dense}, and crystalline~\cite{lorenzo20} states which may, for non-arrested states, give rise also to interesting correlations of the displacements and structural relaxation events.  

In this context, the efficient implementation of activity-driven dynamics (ADD)~\cite{mandal2021how,mandal2020multiple} enables us to study the relaxation of dense systems of self-propelled particles in which the persistence time $\tau_p$ is large compared to the microscopic time $\tau_0$ that the system needs to reach an arrested state for a given set of self-propulsion forces. On time scales $t = t^{\prime}\tau_p$ of the order of the persistence time, the dynamics then becomes intermittent (Fig.~\ref{fig:dep}). In the absence of rearrangements, the system reacts elastically to changes in the self-propulsion forces. The resulting movements are correlated on the length scale of the system (Fig.~\ref{fig:elastic}). Consecutive elastic events may be interrupted by plastic events that trigger instantaneous rearrangements. The participation in these events has a broad distribution; outside of the plastic core forming these avalanches, the remainder of the system moves collectively to accommodate the rearranging regions (Fig.~\ref{fig:plastic}).

Relaxation of the whole structure happens through the accumulation of many of these plastic events. This relaxation dynamics is nearly diffusive at all times (Fig.~\ref{fig:dynf}) and spatially heterogeneous (Fig.~\ref{fig:heter}), implying that plastic events are not independent and tend to concentrate where they have already happened, in a fashion reminiscent of dynamic facilitation.

Dense assemblies of self-propelled particles in two dimensions may serve as a proxy for confluent cell tissues \cite{angelini2011glasslike} in the study of their collective motion \cite{henkes2020dense}. We also expect our results to be transferable from two dimensions to three dimensions, as are the salient features of the physics of glasses \cite{illing2017mermin} and cooperative motion in dense active matter \cite{wysocki2014cooperative}.

The limit of large persistence $\tau_p \to \infty$ is taken at fixed number of particles $N$ and there is thus a dynamical length scale that scales with the system size. As a consequence, the average dynamics of the system and its fluctuations all depend on the system size (Figs.~\ref{fig:dynN},~\ref{fig:heter}).

Our computational abilities do not yet enable us to explore the $f \to 0$ limit where the relaxation time would become large even in rescaled units $t/\tau_p$. At fixed $N$, we expect that the typical time between plastic events $\tau_{\mathrm{res}}$ would become $\gg \tau_p$ in the limit of small $f$. We would then presumably recover a two-step relaxation scenario, with $\beta$-relaxation corresponding to the diffusive elastic exploration of a potential energy minimum, where the localisation length scale may itself depend on the system size. Further explorations are needed to validate this scenario and the nature of the very slow dynamics emerging in this limit. At fixed $f$ we would also expect quantitative changes as $\rho$ is increased: avalanches should become rarer but the evolution of their distribution remains uncertain. Also interesting to study in future will be the connection between plasticity in ADD and approaches to yielding in passive materials that argue in favour of a mechanism based on fluctuating energy barriers~\cite{NicMarBar14} rather than effective thermal activation over fixed barriers~\cite{SolLeqHebCat97,Sollich98,FieSolCat00,PouGut96,BehBiChaHenHar08,RedForPou11}.

\begin{acknowledgments}
We thank E.~Bertin for useful discussions. This work was supported by a grant from the Simons Foundation (\#454933, LB) and by a Visiting Professorship from the Leverhulme Trust (VP1-2019-029, LB). This project has received funding from the European Union’s Horizon 2020 research and innovation programme under Marie Skłodowska-Curie Grant Agreement No.\ 893128 (RM).
\end{acknowledgments}

\bibliographystyle{vancouver}
\bibliography{ref}

\begin{thebibliography}{10}

\bibitem{vicsek95}
Vicsek T, Czir\'ok A, Ben-Jacob E, Cohen I, Shochet O.
\newblock Novel Type of Phase Transition in a System of Self-Driven Particles.
\newblock Phys Rev Lett. 1995 Aug;75:1226-9.
\newblock Available from:
  \url{https://link.aps.org/doi/10.1103/PhysRevLett.75.1226}.

\bibitem{marchetti13}
Marchetti MC, Joanny JF, Ramaswamy S, Liverpool TB, Prost J, Rao M, et~al.
\newblock Hydrodynamics of soft active matter.
\newblock Rev Mod Phys. 2013 Jul;85:1143-89.
\newblock Available from:
  \url{https://link.aps.org/doi/10.1103/RevModPhys.85.1143}.

\bibitem{volpe16}
Bechinger C, Di~Leonardo R, L\"owen H, Reichhardt C, Volpe G, Volpe G.
\newblock Active particles in complex and crowded environments.
\newblock Rev Mod Phys. 2016 Nov;88:045006.
\newblock Available from:
  \url{https://link.aps.org/doi/10.1103/RevModPhys.88.045006}.

\bibitem{solon15}
Solon AP, Fily Y, Baskaran A, Cates ME, Kafri Y, Kardar M, et~al.
\newblock {Pressure is not a state function for generic active fluids}.
\newblock Nature Physics. 2015;11(8):673-8.

\bibitem{berthierkurchan13}
Berthier L, Kurchan J.
\newblock Non-equilibrium glass transitions in driven and active matter.
\newblock Nature Physics. 2013;9(5):310-4.

\bibitem{berthier14}
Berthier L.
\newblock Nonequilibrium Glassy Dynamics of Self-Propelled Hard Disks.
\newblock Phys Rev Lett. 2014 Jun;112:220602.
\newblock Available from:
  \url{https://link.aps.org/doi/10.1103/PhysRevLett.112.220602}.

\bibitem{ni13}
Ni R, Stuart MAC, Dijkstra M.
\newblock Pushing the glass transition towards random close packing using
  self-propelled hard spheres.
\newblock Nat Commun. 2013;4(1):1-7.

\bibitem{szamel2015glassy}
Szamel G, Flenner E, Berthier L.
\newblock Glassy dynamics of athermal self-propelled particles: Computer
  simulations and a nonequilibrium microscopic theory.
\newblock Phys Rev E. 2015 Jun;91:062304.
\newblock Available from:
  \url{https://link.aps.org/doi/10.1103/PhysRevE.91.062304}.

\bibitem{mandal16}
Mandal R, Bhuyan PJ, Rao M, Dasgupta C.
\newblock Active fluidization in dense glassy systems.
\newblock Soft Matter. 2016;12:6268-76.
\newblock Available from: \url{http://dx.doi.org/10.1039/C5SM02950C}.

\bibitem{leomach19}
Klongvessa N, Ginot F, Ybert C, Cottin-Bizonne C, Leocmach M.
\newblock Active Glass: Ergodicity Breaking Dramatically Affects Response to
  Self-Propulsion.
\newblock Phys Rev Lett. 2019 Dec;123:248004.
\newblock Available from:
  \url{https://link.aps.org/doi/10.1103/PhysRevLett.123.248004}.

\bibitem{berthier2019glassy}
Berthier L, Flenner E, Szamel G.
\newblock Glassy Dynamics in Dense Systems of Active Particles.
\newblock The Journal of Chemical Physics. 2019 May;150(20):200901.

\bibitem{janssen2019active}
Janssen LMC.
\newblock Active Glasses.
\newblock Journal of Physics: Condensed Matter. 2019 Dec;31(50):503002.

\bibitem{mandal2020extreme}
Mandal R, Bhuyan PJ, Chaudhuri P, Dasgupta C, Rao M.
\newblock Extreme Active Matter at High Densities.
\newblock Nature Communications. 2020 Dec;11(1):2581.

\bibitem{ManSol21}
Mandal R, Sollich P.
\newblock Shear-Induced Orientational Ordering in an Active Glass Former.
\newblock Proc Natl Acad Sci. 2021 Sep;118(39):e2101964118.
\newblock Available from:
  \url{http://www.pnas.org/lookup/doi/10.1073/pnas.2101964118}.

\bibitem{keta2022disordered}
Keta YE, Jack RL, Berthier L.
\newblock Disordered {{Collective Motion}} in {{Dense Assemblies}} of
  {{Persistent Particles}}.
\newblock Physical Review Letters. 2022 Jul;129(4):048002.

\bibitem{dunkel2013fluid}
Dunkel J, Heidenreich S, Drescher K, Wensink HH, B{\"a}r M, Goldstein RE.
\newblock Fluid {{Dynamics}} of {{Bacterial Turbulence}}.
\newblock Physical Review Letters. 2013 May;110(22):228102.

\bibitem{vicsek2012collective}
Vicsek T, Zafeiris A.
\newblock Collective Motion.
\newblock Physics Reports. 2012 Aug;517(3-4):71-140.

\bibitem{cavagna2018physics}
Cavagna A, Giardina I, Grigera TS.
\newblock The Physics of Flocking: {{Correlation}} as a Compass from
  Experiments to Theory.
\newblock Physics Reports. 2018 Jan;728:1-62.

\bibitem{szamel2014self}
Szamel G.
\newblock Self-propelled particle in an external potential: Existence of an
  effective temperature.
\newblock Phys Rev E. 2014 Jul;90:012111.
\newblock Available from:
  \url{https://link.aps.org/doi/10.1103/PhysRevE.90.012111}.

\bibitem{fodor2016how}
Fodor {\'E}, Nardini C, Cates ME, Tailleur J, Visco P, {van Wijland} F.
\newblock How {{Far}} from {{Equilibrium Is Active Matter}}?
\newblock Physical Review Letters. 2016 Jul;117(3):038103.

\bibitem{mandal2021how}
Mandal R, Sollich P.
\newblock How to Study a Persistent Active Glassy System.
\newblock Journal of Physics: Condensed Matter. 2021 May;33(18):184001.

\bibitem{maloney2006amorphous}
Maloney CE, Lema\^{\i}tre A.
\newblock Amorphous systems in athermal, quasistatic shear.
\newblock Phys Rev E. 2006 Jul;74:016118.
\newblock Available from:
  \url{https://link.aps.org/doi/10.1103/PhysRevE.74.016118}.

\bibitem{flenner2016nonequilibrium}
Flenner E, Szamel G, Berthier L.
\newblock The nonequilibrium glassy dynamics of self-propelled particles.
\newblock Soft matter. 2016;12(34):7136-49.

\bibitem{berthier2017active}
Berthier L, Flenner E, Szamel G.
\newblock How active forces influence nonequilibrium glass transitions.
\newblock New Journal of Physics. 2017;19(12):125006.

\bibitem{henkes2020dense}
Henkes S, Kostanjevec K, Collinson JM, Sknepnek R, Bertin E.
\newblock Dense Active Matter Model of Motion Patterns in Confluent Cell
  Monolayers.
\newblock Nature Communications. 2020 Dec;11(1):1405.

\bibitem{szamel2021long}
Szamel G, Flenner E.
\newblock Long-ranged velocity correlations in dense systems of self-propelled
  particles.
\newblock Europhysics Letters. 2021;133(6):60002.

\bibitem{mandal2020multiple}
Mandal R, Sollich P.
\newblock Multiple Types of Aging in Active Glasses.
\newblock Physical Review Letters. 2020;125(21):218001.

\bibitem{liao2018criticality}
Liao Q, Xu N.
\newblock Criticality of the Zero-Temperature Jamming Transition Probed by
  Self-Propelled Particles.
\newblock Soft Matter. 2018;14(5):853-60.

\bibitem{villarroel2021critical}
Villarroel C, D{\"u}ring G.
\newblock Critical Yielding Rheology: From Externally Deformed Glasses to
  Active Systems.
\newblock Soft Matter. 2021;17(43):9944-9.

\bibitem{morse2021direct}
Morse PK, Roy S, Agoritsas E, Stanifer E, Corwin EI, Manning ML.
\newblock A Direct Link between Active Matter and Sheared Granular Systems.
\newblock Proceedings of the National Academy of Sciences. 2021 May;118(18).

\bibitem{sethna1993hysteresis}
Sethna JP, Dahmen K, Kartha S, Krumhansl JA, Roberts BW, Shore JD.
\newblock Hysteresis and hierarchies: Dynamics of disorder-driven first-order
  phase transformations.
\newblock Phys Rev Lett. 1993 May;70:3347-50.

\bibitem{liu2016driving}
Liu C, Ferrero EE, Puosi F, Barrat JL, Martens K.
\newblock Driving {{Rate Dependence}} of {{Avalanche Statistics}} and
  {{Shapes}} at the {{Yielding Transition}}.
\newblock Physical Review Letters. 2016 Feb;116(6):065501.

\bibitem{ozawa2018random}
Ozawa M, Berthier L, Biroli G, Rosso A, Tarjus G.
\newblock Random Critical Point Separates Brittle and Ductile Yielding
  Transitions in Amorphous Materials.
\newblock Proceedings of the National Academy of Sciences. 2018
  Jun;115(26):6656-61.

\bibitem{angelini2011glasslike}
Angelini TE, Hannezo E, Trepat X, Marquez M, Fredberg JJ, Weitz DA.
\newblock Glass-like Dynamics of Collective Cell Migration.
\newblock Proceedings of the National Academy of Sciences. 2011
  Mar;108(12):4714-9.

\bibitem{kumar2014flocking}
Kumar N, Soni H, Ramaswamy S, Sood AK.
\newblock Flocking at a distance in active granular matter.
\newblock Nature Communications. 2014;5(1):4688.

\bibitem{dauchot2016crystal}
Briand G, Dauchot O.
\newblock Crystallization of Self-Propelled Hard Discs.
\newblock Phys Rev Lett. 2016 Aug;117:098004.

\bibitem{berthier2011dynamical}
Berthier L, Biroli G, Bouchaud JP, Cipelletti L, van Saarloos W.
\newblock Dynamical heterogeneities in glasses, colloids, and granular media.
  vol. 150.
\newblock OUP Oxford; 2011.

\bibitem{chandler2011}
Keys AS, Hedges LO, Garrahan JP, Glotzer SC, Chandler D.
\newblock Excitations Are Localized and Relaxation Is Hierarchical in
  Glass-Forming Liquids.
\newblock Phys Rev X. 2011 Nov;1:021013.
\newblock Available from:
  \url{https://link.aps.org/doi/10.1103/PhysRevX.1.021013}.

\bibitem{berthierbiroli11}
Berthier L, Biroli G.
\newblock Theoretical perspective on the glass transition and amorphous
  materials.
\newblock Rev Mod Phys. 2011 Jun;83:587-645.
\newblock Available from:
  \url{https://link.aps.org/doi/10.1103/RevModPhys.83.587}.

\bibitem{fily2014freezing}
Fily Y, Henkes S, Marchetti MC.
\newblock Freezing and Phase Separation of Self-Propelled Disks.
\newblock Soft Matter. 2014;10(13):2132-40.

\bibitem{digregorio2018full}
Digregorio P, Levis D, Suma A, Cugliandolo LF, Gonnella G, Pagonabarraga I.
\newblock Full {{Phase Diagram}} of {{Active Brownian Disks}}: {{From Melting}}
  to {{Motility-Induced Phase Separation}}.
\newblock Physical Review Letters. 2018 Aug;121(9):098003.

\bibitem{alglib}
ALGLIB. Unconstrained optimization: {L-BFGS} and {CG}; 2021.
\newblock Accessed: 2021-04-02.
\newblock Available from:
  \url{https://www.alglib.net/optimization/lbfgsandcg.php}.

\bibitem{nishikawa2022relaxation}
Nishikawa Y, Ozawa M, Ikeda A, Chaudhuri P, Berthier L.
\newblock Relaxation {{Dynamics}} in the {{Energy Landscape}} of
  {{Glass-Forming Liquids}}.
\newblock Physical Review X. 2022 Apr;12(2):021001.

\bibitem{maloney2006correlations}
Maloney CE.
\newblock Correlations in the {{Elastic Response}} of {{Dense Random
  Packings}}.
\newblock Physical Review Letters. 2006 Jul;97(3):035503.

\bibitem{slaughter2002linearized}
Slaughter WS.
\newblock The {{Linearized Theory}} of {{Elasticity}}.
\newblock {Boston, MA}: {Birkh\"auser Boston}; 2002.

\bibitem{mizuno2017continuum}
Mizuno H, Shiba H, Ikeda A.
\newblock Continuum Limit of the Vibrational Properties of Amorphous Solids.
\newblock Proceedings of the National Academy of Sciences. 2017 Nov;114(46).

\bibitem{lemaitre2014structural}
Lema{\^i}tre A.
\newblock Structural {{Relaxation}} Is a {{Scale-Free Process}}.
\newblock Physical Review Letters. 2014 Dec;113(24):245702.

\bibitem{lerbinger2021relevance}
Lerbinger M, Barbot A, Vandembroucq D, Patinet S.
\newblock Relevance of Shear Transformations in the Relaxation of Supercooled
  Liquids.
\newblock Phys Rev Lett. 2022 Oct;129:195501.
\newblock Available from:
  \url{https://link.aps.org/doi/10.1103/PhysRevLett.129.195501}.

\bibitem{karimi2017inertia}
Karimi K, Ferrero EE, Barrat JL.
\newblock Inertia and Universality of Avalanche Statistics: {{The}} Case of
  Slowly Deformed Amorphous Solids.
\newblock Physical Review E. 2017 Jan;95(1):013003.

\bibitem{maloney2004subextensive}
Maloney C, Lema{\^i}tre A.
\newblock Subextensive {{Scaling}} in the {{Athermal}}, {{Quasistatic Limit}}
  of {{Amorphous Matter}} in {{Plastic Shear Flow}}.
\newblock Physical Review Letters. 2004 Jul;93(1):016001.

\bibitem{maloneyprivate}
Maloney C; 2022.
\newblock Private conversation.

\bibitem{binder2005glassy}
Binder K, Kob W.
\newblock Glassy {{Materials}} and {{Disordered Solids}}: {{An Introduction}}
  to {{Their Statistical Mechanics}}.
\newblock {World Scientific Publishing Company}; 2005.

\bibitem{heussinger2010superdiffusive}
Heussinger C, Berthier L, Barrat JL.
\newblock Superdiffusive, heterogeneous, and collective particle motion near
  the fluid-solid transition in athermal disordered materials.
\newblock EPL (Europhysics Letters). 2010;90(2):20005.

\bibitem{kob1997dynamical}
Kob W, Donati C, Plimpton SJ, Poole PH, Glotzer SC.
\newblock Dynamical {{Heterogeneities}} in a {{Supercooled Lennard-Jones
  Liquid}}.
\newblock Physical Review Letters. 1997 Oct;79(15):2827-30.

\bibitem{chaudhuri2007universal}
Chaudhuri P, Berthier L, Kob W.
\newblock Universal Nature of Particle Displacements close to Glass and Jamming
  Transitions.
\newblock Phys Rev Lett. 2007 Aug;99:060604.
\newblock Available from:
  \url{https://link.aps.org/doi/10.1103/PhysRevLett.99.060604}.

\bibitem{berthier2005length}
Berthier L, Chandler D, Garrahan JP.
\newblock Length Scale for the Onset of {{Fickian}} Diffusion in Supercooled
  Liquids.
\newblock Europhysics Letters (EPL). 2005 Feb;69(3):320-6.

\bibitem{berthier2004time}
Berthier L.
\newblock Time and Length Scales in Supercooled Liquids.
\newblock Physical Review E. 2004 Feb;69(2):020201.

\bibitem{henkes2012extracting}
Henkes S, Brito C, Dauchot O.
\newblock Extracting Vibrational Modes from Fluctuations: A Pedagogical
  Discussion.
\newblock Soft Matter. 2012;8(22):6092.

\bibitem{bottinelli2017how}
Bottinelli A, Silverberg JL.
\newblock How to: {{Using Mode Analysis}} to {{Quantify}}, {{Analyze}}, and
  {{Interpret}} the {{Mechanisms}} of {{High-Density Collective Motion}}.
\newblock Frontiers in Applied Mathematics and Statistics. 2017 Dec;3:26.

\bibitem{pal2022inspection}
Pal A, Kostinski S, Reuveni S.
\newblock The Inspection Paradox in Stochastic Resetting.
\newblock Journal of Physics A: Mathematical and Theoretical. 2022
  Jan;55(2):021001.

\bibitem{shiba2012relationship}
Shiba H, Kawasaki T, Onuki A.
\newblock Relationship between Bond-Breakage Correlations and Four-Point
  Correlations in Heterogeneous Glassy Dynamics: {{Configuration}} Changes and
  Vibration Modes.
\newblock Physical Review E. 2012 Oct;86(4):041504.

\bibitem{lemaitre2007plastic}
Lema\^{\i}tre A, Caroli C.
\newblock Plastic response of a two-dimensional amorphous solid to quasistatic
  shear: Transverse particle diffusion and phenomenology of dissipative events.
\newblock Phys Rev E. 2007 Sep;76:036104.
\newblock Available from:
  \url{https://link.aps.org/doi/10.1103/PhysRevE.76.036104}.

\bibitem{guiselin2022microscopic}
Guiselin B, Scalliet C, Berthier L.
\newblock Microscopic origin of excess wings in relaxation spectra of
  supercooled liquids.
\newblock Nature Physics. 2022;18(4):468-72.

\bibitem{lacevic2003spatially}
La{\v c}evi{\'c} N, Starr FW, Schr{\o}der TB, Glotzer SC.
\newblock Spatially Heterogeneous Dynamics Investigated via a Time-Dependent
  Four-Point Density Correlation Function.
\newblock The Journal of Chemical Physics. 2003 Oct;119(14):7372-87.

\bibitem{scalliet2022thirty}
Scalliet C, Guiselin B, Berthier L.
\newblock Thirty Milliseconds in the Life of a Supercooled Liquid.
\newblock Phys Rev X. 2022 Dec;12:041028.
\newblock Available from:
  \url{https://link.aps.org/doi/10.1103/PhysRevX.12.041028}.

\bibitem{PhysRevX.1.021013}
Keys AS, Hedges LO, Garrahan JP, Glotzer SC, Chandler D.
\newblock Excitations Are Localized and Relaxation Is Hierarchical in
  Glass-Forming Liquids.
\newblock Phys Rev X. 2011 Nov;1:021013.
\newblock Available from:
  \url{https://link.aps.org/doi/10.1103/PhysRevX.1.021013}.

\bibitem{lorenzo20}
Caprini L, Marini Bettolo~Marconi U, Puglisi A.
\newblock Spontaneous Velocity Alignment in Motility-Induced Phase Separation.
\newblock Phys Rev Lett. 2020 Feb;124:078001.
\newblock Available from:
  \url{https://link.aps.org/doi/10.1103/PhysRevLett.124.078001}.

\bibitem{illing2017mermin}
Illing B, Fritschi S, Kaiser H, Klix CL, Maret G, Keim P.
\newblock Mermin\textendash{{Wagner}} Fluctuations in {{2D}} Amorphous Solids.
\newblock Proceedings of the National Academy of Sciences. 2017
  Feb;114(8):1856-61.

\bibitem{wysocki2014cooperative}
Wysocki A, Winkler RG, Gompper G.
\newblock Cooperative Motion of Active {{Brownian}} Spheres in
  Three-Dimensional Dense Suspensions.
\newblock EPL (Europhysics Letters). 2014 Feb;105(4):48004.

\bibitem{NicMarBar14}
Nicolas A, Martens K, Barrat JL.
\newblock Rheology of Athermal Amorphous Solids: Revisiting Simplified
  Scenarios and the Concept of Mechanical Noise Temperature.
\newblock Europhys Lett. 2014;107:44003.

\bibitem{SolLeqHebCat97}
Sollich P, Lequeux F, H{\'e}braud P, Cates ME.
\newblock Rheology of Soft Glassy Materials.
\newblock Phys Rev Lett. 1997;78:2020-3.

\bibitem{Sollich98}
Sollich P.
\newblock Rheological Constitutive Equation for a Model of Soft Glassy
  Materials.
\newblock Phys Rev E. 1998;58:738-59.

\bibitem{FieSolCat00}
Fielding SM, Sollich P, Cates ME.
\newblock Aging and Rheology in Soft Materials.
\newblock J Rheol. 2000;44(2):323-69.

\bibitem{PouGut96}
Pouliquen O, Gutfraind R.
\newblock Stress Fluctuations and Shear Zones in Quasistatic Granular Chute
  Flows.
\newblock Phys Rev E. 1996 Jan;53(1):552-61.
\newblock Available from:
  \url{https://link.aps.org/doi/10.1103/PhysRevE.53.552}.

\bibitem{BehBiChaHenHar08}
Behringer RP, Bi D, Chakraborty B, Henkes S, Hartley RR.
\newblock Why {{Do Granular Materials Stiffen}} with {{Shear Rate}}? {{Test}}
  of {{Novel Stress-Based Statistics}}.
\newblock Phys Rev Lett. 2008 Dec;101(26):268301.
\newblock Available from:
  \url{https://link.aps.org/doi/10.1103/PhysRevLett.101.268301}.

\bibitem{RedForPou11}
Reddy KA, Forterre Y, Pouliquen O.
\newblock Evidence of {{Mechanically Activated Processes}} in {{Slow Granular
  Flows}}.
\newblock Phys Rev Lett. 2011 Mar;106(10):108301.
\newblock Available from:
  \url{https://link.aps.org/doi/10.1103/PhysRevLett.106.108301}.

\end{thebibliography}

\end{document}